\documentclass{elsart}
\usepackage{amssymb}
\newtheorem{lemma}{Lemma}
\newtheorem{theorem}{Theorem}

\begin{document}

\begin{frontmatter}
\title{A lower bound on the 2-adic complexity of Ding-Helleseth generalized cyclotomic sequence of period $p^n$ }
\author{Yuhua Sun$^{a,b,c}$, Qiang Wang$^{b}$, Tongjiang Yan$^{a,c}$,Chun'e Zhao$^{a,c}$}
\thanks{The work is supported by Shandong Provincial Natural Science Foundation of China (No. ZR2014FQ005), Fundamental Research Funds for the Central Universities (No. 15CX02065A, No. 15CX08011A, No. 15CX02056A, No. 16CX02013A, No. 16CX02009A), Fujian Provincial Key Laboratory of Network Security and Cryptology Research Fund (Fujian Normal University) (No.15002),Qingdao application research on special independent innovation plan project (No. 16-5-1-5-jch)}
\address[label1]{College of Sciences,
China University of Petroleum,
Qingdao 266555,
Shandong, China(e-mail:sunyuhua\_1@163.com), (e-mail:wang@math.carleton.ca), (e-mail:yantoji@163.com), (e-mail:zhaochune1981@163.com).}
\address[label2]{School of Mathematics and Statistics,
Carleton University,
Ottawa,
Ontario, K1S 5B6, Canada.}
\address[label3]{Key Laboratory of Network Security and Cryptology,
Fujian Normal University,
Fuzhou, Fujian 350117,
China}
\begin{abstract}
Let $p$ be an odd prime, $n$ a positive integer and $g$ a primitive root of $p^n$. Suppose $D_i^{(p^n)}=\{g^{2s+i}|s=0,1,2,\cdots,\frac{(p-1)p^{n-1}}{2}\}$, $i=0,1$, is the generalized cyclotomic classes with $Z_{p^n}^{\ast}=D_0\cup D_1$. In this paper, we prove that Gauss periods based on $D_0$ and $D_1$ are both equal to 0 for $n\geq2$. As an application, we determine a lower bound on the 2-adic complexity of a class of Ding-Helleseth generalized cyclotomic sequences of period $p^n$. The result shows that the 2-adic complexity is at least $p^n-p^{n-1}-1$, which is larger than $\frac{N+1}{2}$, where $N=p^n$ is the period of the sequence.
\end{abstract}
\end{frontmatter}
\section{Introduction}
Since Berlekamp-Massey algorithm (BMA) for LFSRs \cite{Massey} and Rational Approximation algorithm for FCSRs \cite{Andrew Klapper} were presented, linear complexity and 2-adic complexity of pseudo-random sequence have been thought as critical security criteria and a sequence is called good sequence if its linear complexity and 2-adic complexity are both larger than one half of its period.

After the security criteria of linear complexity has been given, many sequences families have been constructed and proved to be with high linear complexity (See \cite{Ding Cunsheng}-\cite{Edemskiy}). However, there are only a few kinds of sequences whose 2-adic complexities have been completely determined though the concept of 2-adic complexity has been presented for more than two decades. For instance, Tian and Qi pointed out that all the binary $m$-sequences have maximal 2-adic complexity \cite{Tian Tian}. In 2014, Xiong et al. \cite{Xiong Hai} presented a general method based on circulant matrices to determine the 2-adic complexity of binary sequence. Using this method, they have proved that the 2-adic complexities of all the known sequences with ideal 2-level autocorrelation are maximal. Moreover, they proved that other four classes of sequences with optimal autocorrelation, i.e., Legendre sequences, Ding-Helleseth-Lam sequences \cite{Ding-H-L}, Tang-Ding sequences \cite{Tang-Ding}, and sequences constructed by Zhou et al.\cite{Zhou}, have maximal 2-adic complexity, too.

Sequences from generalized cyclotomic classes are large and important sequence families. Since Ding and Helleseth \cite{Ding-Helleseth} presented some new cyclotomy and pointed out its applications in constructing sequences, many generalized cyclotomic sequences have been constructed and have been prove to have high linear complexities (See \cite{Ding Cunsheng}, \cite{Bai Enjian}, \cite{Yan Tongjiang-1}, \cite{Hu Liqin}, \cite{Li Xiaoping}, \cite{Wang Qiu}, \cite{Yan-thesis}). But most of the 2-adic complexities of these sequences are still unknown.

In this paper, we will compute the Gauss periods based on Ding-Helleseth generalized cyclotomic classes from a prime power $p^n$. As an application, combining the method by Xiong et al. \cite{Xiong Hai}, we study the 2-adic complexity of Ding-Helleseth generalized cyclotomic sequence with period $p^n$ and a lower bound on the 2-adic complexity will be given, i.e., we prove that the 2-adic complexity is at least $p^n-p^{n-1}$, which is larger than $\frac{p^n+1}{2}$, one half of the period of the sequence.

The rest of this paper is organized as follows. In section 2, we will give the necessary definitions, notations, and some previous results. Gauss periods based on Ding-Hellesth generalized cyclotomic classes of order 2 from a prime power $p^n$ will be given in section 3. And a lower bound on the 2-adic complexity of Ding-Helleseth generalized cyclotomic sequence of period $p^n$ will be given in Section 4. Finally we give a summarize on our results in Section 5.

\section{Preliminaries}\label{section 2}
Let $\{s_i\}_{i=0}^{N-1}$ be a binary sequence of period $N$, where $N$ is a positive integer. Suppose $S(x)=\sum\limits_{i=0}^{N-1}s_{i}x^i\in \mathbb{Z} [x]$ and
\begin{equation}
\frac{S(2)}{2^N-1}=\frac{\sum\limits_{i=0}^{N-1}s_{i}2^i}{2^N-1}=\frac{m}{n},\ 0\leq m\leq n,\ \mathrm{gcd}(m,n)=1.\label{2-adic complexity}
\end{equation}
Then the 2-adic complexity $\Phi_{2}(s)$ of the sequence $\{s_i\}_{i=0}^{N-1}$ is defined as the integer $\lfloor\mathrm{log}_2n\rfloor$, i.e.,
\begin{equation}
\Phi_{2}(s)=\left\lfloor\mathrm{log}_2\frac{2^N-1}{\mathrm{gcd}(2^N-1,S(2))}\right\rfloor,\label{2-adic calculation}
\end{equation}
where $\lfloor x\rfloor$ is the greatest integer that is less than or equal to $x$.

Let $p$ be an odd prime and $n$ a positive integer. Then there exists a primitive root $g$ of $p^n$.
Define
\begin{eqnarray}
D_i^{(p^n)}&=&\{g^{2t+i}\mid t=0, 1, \cdots, \frac{(p-1)p^{n-1}}{2}-1\},\ i=0, 1, \nonumber\\
R^{(p^n)}&=&\{0,p,2p,\cdots,(p^{n-1}-1)p\},\nonumber
\end{eqnarray}
then the set $D_i^{(p^n)}$, $i=0,1$, is called Ding-Helleseth generalized cyclotomic class of order 2 from the prime power $p^n$, which was introduced by Ding and Helleseth in \cite{Ding-Helleseth}, and  it is easy to see $\mathbb{Z}_{p^n}=R^{(p^n)}\cup D_0^{(p^n)}\cup D_1^{(p^n)}$. Correspondingly, the generalized cyclotomic numbers is defined as
\begin{eqnarray}
(i,j)^{(p^n)}=|(D_i^{(p^n)}+1)\cap D_j^{(p^n)}|,\ \ i,j=0,1.\nonumber
\end{eqnarray}
For the above sets, Ding and Helleseth \cite{Ding-Helleseth} have also given the following three results, which will be useful for the proof of our main result.
\begin{lemma}
\begin{eqnarray}
\mid R^{(p^n)}\cap (D_1^{(p^n)}+1)\mid&=&\left\{
\begin{array}{ll}
0,\ \ \ \ \  p\equiv1\pmod{4},\\
p^{n-1},\ p\equiv3\pmod{4}.
\end{array}
\right.\nonumber\\
\mid R^{(p^n)}\cap (D_0^{(p^n)}+1)\mid &=&\left\{
\begin{array}{ll}
p^{n-1},\ p\equiv1\pmod{4},\\
0,\ \ \ \ \  p\equiv3\pmod{4}.
\end{array}
\right.\nonumber
\end{eqnarray}
\end{lemma}
\begin{lemma}\label{cyclotomic number}
If $p\equiv3\pmod{4}$, then
$$(1,0)^{(p^n)}=(0,0)^{(p^n)}=(1,1)^{(p^n)}=\frac{p^{n-1}(p-3)}{4},\ (0,1)^{(p^n)}=\frac{p^{n-1}(p+1)}{4}.$$
If $p\equiv1\pmod{4}$, then
$$(0,1)^{(p^n)}=(1,0)^{(p^n)}=(1,1)^{(p^n)}=\frac{p^{n-1}(p-1)}{4},\ (0,0)^{(p^n)}=\frac{p^{n-1}(p-5)}{4}.$$
\end{lemma}
\begin{lemma}\label{cyclotomic equation}
For any $r\in R^{(p^n)}$,
$$
\mid(D_{i}^{(p^n)}+r)\cap D_{j}^{(p^n)}\mid=\left\{
\begin{array}{ll}
p^{n-1}(p-1)/2,\ \mathrm{if}\ i=j,\\
0,\ \ \ \ \ \ \ \mathrm{otherwise}.
\end{array}
\right.
$$
\end{lemma}
Moreover, it is not difficult to get the following result.
\begin{lemma}\label{basic cyclotomic-1}
(1) For $a\in D_{i}^{(p^n)}$, $aD_{j}^{(p^n)}=D_{(i+j)\ (\mathrm{mod}\ 2)}^{(p^n)}$, where $i,j=0,1$;\\
(2) For any $n\geq2$ and two positive integers $n_1,n_2$ with $n_1<n_2\leq n$, the element $x\ (\mathrm{mod}\ p^{n_1})$ runs exactly through every element $p^{n_2-n_1}$ times in $D_i^{(p^{n_1})}$ when $x$ runs through every element in $D_i^{(p^{n_2})}$, and the element $x\ (\mathrm{mod}\ p^{n_1})$ runs exactly through every element $p^{n_2-n_1}$ times in $R^{(p^{n_1})}$ when $x$ runs through every element in $R^{(p^{n_2})}$, where $i=0,1$.
\end{lemma}
According to this property, we can get the following decomposition of $\mathbb{Z}_{p^n}$
\begin{eqnarray}
\mathbb{Z}_{p^n}&=&D_0^{(p^n)}\cup D_1^{(p^n)}\cup p\mathbb{Z}_{p^{n-1}}\nonumber\\
&=&(D_0^{(p^n)}\cup D_0^{(p^{n-1})})\cup(D_1^{(p^n)}\cup D_1^{(p^{n-1})})\cup p^2\mathbb{Z}_{p^{n-2}}\nonumber\\
&&\cdots\nonumber\\
&=&(\cup_{m=1}^{n}p^{n-m}D_0^{(p^m)})\cup(\cup_{m=1}^{n}p^{n-m}D_1^{(p^m)})\cup \{0\}.\label{decomposition}
\end{eqnarray}
Let $C_0=\cup_{m=1}^{n}p^{n-m}D_0^{(p^m)}$, $C_1=\cup_{m=1}^{n}p^{n-m}D_1^{(p^m)}\cup\{0\}$. Then $\mathbb{Z}_{p^n}=C_0\cup C_1$ and $C_0\cap C_1=\emptyset$. The Ding-Helleseth generalized sequence $\{s_i\}_{i=0}^{p^n-1}$ of period $p^n$ is defined by
\begin{eqnarray}
s_i=\left\{
\begin{array}{ll}
0,\ \mathrm{if}\ i\pmod{p^n}\in C_0,\\
1,\ \mathrm{if}\ i\pmod{p^n}\in C_1.
\end{array}
\right.\label{sequence defi}
\end{eqnarray}

Let $\omega_{p^n}=e^{2\pi\sqrt{-1}/p^n}$ be a $p^n$th complex primitive root of unity. Then the additive character $\chi$ of $\mathbb{Z}_{p^n}$ is given by
$$
\chi(x)=\omega_{p^n}^{x},\ x\in \mathbb{Z}_{p^n}
$$
and Gaussian periods are defined by
\begin{equation}
\eta_{i}^{(p^n)}=\sum_{x\in D_{i}^{(p^n)}}\chi(x),\ i=0,1.\label{gauss period}
\end{equation}
In this paper, we will prove that the gaussian periods in Eq. (\ref{gauss period}) are both 0 for any $n\geq2$. Then, using the gaussian periods, a lower bound on the 2-adic of Ding-Helleseth generalized cyclotomic sequence of period $p^n$ will be derived. And the result shows that the 2-adic complexity is large enough to resist the RAA for FCSRs.
\section{Gaussian periods of Ding-Hellesth generalized cyclotomic classes from the prime power $p^n$}
In this section, we will determine the Gaussian periods defined in Eq. (\ref{gauss period}). In order to do this, we need the following result which can be easily obtained, and we omitted the proof.
\begin{lemma}\label{n equal one}
With the notations as above, then for $n=1$ the gauss periods satisfy
$\eta_0^{(p)}+\eta_1^{(p)}+1=0$, and for $n\geq2$ we have $\sum_{x\in R^{(p^n)}}\chi(x)=0.$
\end{lemma}
We will also use the following Lemma, which has been proved by Store \cite{store}.
\begin{lemma}
 If $p\equiv1\pmod{4}$, then $-1\in D_0^{(p^n)}$. If $p\equiv3\pmod{4}$, then $-1\in D_1^{(p^n)}$.
\end{lemma}
Now, we give our first main result.
\begin{theorem}\label{gauss period-1}
Let $p$ be an odd prime, $n\geq2$ positive integer and $D_0^{(p^n)},D_1^{(p^n)}$ the Ding-Hellesth generalized cyclotomic classes of ode 2 from the prime power $p^n$. Suppose that $\eta_0^{(p^n)}$ and $\eta_1^{(p^n)}$ are the Gauss periods based on $D_0^{(p^n)}$ and $D_1^{(p^n)}$. Then we have $\eta_0^{(p^n)}=\eta_1^{(p^n)}=0$.
\end{theorem}
\noindent{\bf Proof.} Note that $\sum_{x\in \mathbb{Z}_{p^n}}\omega_{p^n}^x=0$. By Eq. (\ref{decomposition}), we have
\begin{eqnarray}
\sum_{x\in \mathbb{Z}_{p^n}}\omega_{p^n}^x&=&1+\sum_{m=1}^{n}\sum_{x\in p^{n-m}D_0^{(p^m)}}\omega_{p^n}^{x}+\sum_{m=1}^{n}\sum_{x\in p^{n-m}D_1^{(p^m)}}\omega_{p^n}^{x}\nonumber\\
&=&1+\sum_{m=1}^{n}\sum_{x\in D_0^{(p^m)}}\omega_{p^m}^{x}+\sum_{m=1}^{n}\sum_{x\in D_1^{(p^m)}}\omega_{p^m}^{x}\nonumber\\
&=&1+\sum_{m=1}^{n}(\eta_0^{(p^m)}+\eta_1^{(p^m)})\nonumber\\
&=&1+\eta_0^{(p)}+\eta_1^{(p)}+\sum_{m=2}^{n}(\eta_0^{(p^m)}+\eta_1^{(p^m)})\nonumber\\
&=&0.
\end{eqnarray}
By Lemma \ref{n equal one}, we derive
$$\sum_{m=2}^{n}(\eta_0^{(p^m)}+\eta_1^{(p^m)})=0.$$
Take $n=2$, we get $\eta_0^{(p^2)}+\eta_1^{(p^2)}=0$. Using induction, it is not difficult to get
\begin{equation}
\eta_0^{(p^n)}+\eta_1^{(p^n)}=0\label{m case}
\end{equation}
for all $n\geq2$. Furthermore, if $p\equiv1\pmod{4}$, then $-1\in D_0^{(p^n)}$ and $(0,1)^{(p^n)}=(1,0)^{(p^n)}$. So, by Lemmas \ref{cyclotomic number} and \ref{cyclotomic equation}, for $n\geq2$, we have
\begin{eqnarray}
\eta_0^{(p^n)}\eta_1^{(p^n)}&=&\left(\sum_{x\in D_0^{(p^n)}}\omega_{p^n}^{x}\right)\left(\sum_{y\in D_1^{(p^n)}}\omega_{p^n}^{x}\right)\nonumber\\
&=&\sum_{x\in D_0^{(p^n)}}\sum_{y\in D_1^{(p^n)}}\omega_{p^n}^{y-x}\nonumber\\
&=&(0,1)^{(p^n)}\eta_0^{(p^n)}+(1,0)^{(p^n)}\eta_1^{(p^n)}+0\nonumber\\
&=&(0,1)^{(p^n)}(\eta_0^{(p^n)}+\eta_1^{(p^n)})\nonumber\\
&=&0,\nonumber
\end{eqnarray}
where the last equation comes from Eq. (\ref{m case}). And again by Eq. (\ref{m case}), we get the desire result.
For the case of $p\equiv3\pmod{4}$, we can also get the similar result except that it additionally need the result $\sum_{x\in R^{(p^n)}}\chi(x)=0$ in Lemma \ref{n equal one}.\ \ \ \ \ \ \ \ \ \ \ \ \ \ \ \ \ \ \ \ \ \ \ \ \ \ \ \ \ \ \ \ \ \ \ \ \ \ \ \ \ \ \ \ \ \ \ \ \ \ \ \ \ \ \ \ \ \ \ \ \ \ \ \ \ \ \ \ \ \ \ \ \ \ \ \ \ \ \ \ \ \ \ \ \ \ \ \ \ \ \ \ \ \ \ \ \ \ \ \ \ \ \ \ \ $\Box$
\section{Lower bound on the 2-adic complexity of Ding-Helleseth generalized cyclotomic sequence of period $p^n$} \label{section 4}
Let $p$ be an odd prime, $n$ a positive integer, $N=p^n$, and $\{s_i\}_{i=0}^{N-1}$ the Ding-Helleseth generalized cyclotomic sequence of period $N$ defined in Eq. (\ref{sequence defi}). Suppose $A=(a_{i,j})_{N\times N}$ is a matrix of order $N$, where $a_{i,j}=s_{i-j\pmod{N}}$. In this section, inspired by the method of Xiong et al. \cite{Xiong Hai}, we will give a lower bound on the 2-adic complexity of $\{s_i\}_{i=0}^{N-1}$. To this end, we will need the following three results, which can be found in \cite{Xiong Hai}, \cite{Davis}, and \cite{store} respectively.
\begin{lemma}\label{main method-1}\cite{Xiong Hai}
Let $A$ be regarded  as a matrix over $\mathbb{Q}$, the rational field. Suppose $\mathrm{det}(A)\neq0$. Then
\begin{equation}
\mathrm{gcd}\left(S(2),2^N-1\right)|\mathrm{gcd}\left(\mathrm{det}(A),2^N-1\right).\label{method-1}
\end{equation}
\end{lemma}
\begin{lemma}\label{main method-2}\cite{Davis}
Let $\{s_i\}_{i=0}^{N-1}$ be a binary sequence of period $N$ and $A=(a_{k,j})_{N\times N}$ the circulant matrix determined by $a_{k,j}=s_{(k-j)\pmod{N}}$. Then $\mathrm{det}(A)=\prod_{a=0}^{N-1}S(\omega_N^a)$, where $\omega_N$ is the $N$th complex primitive root of  unit.
\end{lemma}
\begin{lemma}\label{prod}
\cite{store} Let the symbols be the same as above. Then $\eta_0^{(p)}\eta_1^{(p)}=\frac{1-p}{4}$ for $p\equiv1\ (\mathrm{mod}\ 4)$ and $\eta_0^{(p)}\eta_1^{(p)}=\frac{1+p}{4}$ for $p\equiv3\ (\mathrm{mod}\ 4)$.
\end{lemma}
Additionally, we also need the following two Lemmas.
\begin{lemma}\label{main method-3}
Let $\{s_i\}_{i=0}^{N-1}$ be the Ding-Helleseth generalized cyclotomic sequence of period $p^n$. Then we have
\begin{equation}
S(\omega_{N}^{a})=\left\{
\begin{array}{lllll}
\frac{p^n+1}{2},\ \ \ \ \ \ \ \ \ \ \ \ \ \mathrm{if}\ a=0,\\
\frac{p^m+1}{2}+p^{m}\eta_1^{(p)},\ \mathrm{if}\ a\in p^{m}D_0^{(p^{n-m})},\ m=0,1,2,\cdots,n-1.\\
\frac{p^m+1}{2}+p^{m}\eta_0^{(p)},\ \mathrm{if}\ a\in p^{m}D_1^{(p^{n-m})},\ m=0,1,2,\cdots,n-1.
\end{array}
\right.
\end{equation}
\end{lemma}
\noindent{\bf Proof.}
Note that
\begin{eqnarray}
S(\omega_N)=1+\Sigma_{k=1}^{n}\sum_{x\in p^{n-k}D_1^{(p^k)}}\omega_N^x\nonumber
\end{eqnarray}
For $a=0$, it is obvious that $S(\omega_N^a)=\frac{p^n+1}{2}$.

For $0\leq m\leq n-1$ and $a\in p^mD_0^{(p^{n-m})}$, by Lemma \ref{basic cyclotomic-1}, we have $\sum_{x\in p^{n-k}D_1^{(p^k)}}\omega_N^{ax}=\frac{p^{k-1}(p-1)}{2}$ if $k\leq m$ and $\sum_{x\in p^{n-k}D_1^{(p^k)}}\omega_N^{ax}=p^m\sum_{x\in p^{n-k+m}D_1^{(p^{k-m})}}\omega_N^{x}$ if $m<k\leq n$, i.e.,
\begin{eqnarray}
S(\omega_N^a)&=&1+\sum_{k=1}^{m}\sum_{x\in p^{n-k}D_1^{(p^k)}}\omega_N^{ax}+\sum_{k=m+1}^{n}\sum_{x\in p^{n-k}D_1^{(p^k)}}\omega_N^{ax}\nonumber\\
&=&1+\sum_{k=1}^{m}\frac{p^{k-1}(p-1)}{2}+\sum_{k=m+1}^{n}p^m\sum_{x\in p^{n-k+m}D_1^{(p^{k-m})}}\omega_N^{x}\nonumber\\
&=&1+\frac{p^m-1}{2}+p^m\sum_{k=m+1}^{n}\eta_1^{(p^{k-m})}\nonumber\\
&=&1+\frac{p^m-1}{2}+p^m\eta_1^{(p)}=\frac{p^m+1}{2}+p^m\eta_1^{(p)},\nonumber
\end{eqnarray}
where the last equation is by Theorem \ref{gauss period-1}, $\eta_1^{(p^i)}=0$ for $i\geq2$.

Similarly, for $0\leq m\leq n-1$ and $a\in p^mD_1^{(p^{n-m})}$, we have
\begin{eqnarray}
S(\omega_N^a)&=&1+\sum_{k=1}^{m}\sum_{x\in p^{n-k}D_1^{(p^k)}}\omega_N^{ax}+\sum_{k=m+1}^{n}\sum_{x\in p^{n-k}D_1^{(p^k)}}\omega_N^{ax}\nonumber\\
&=&1+\sum_{k=1}^{m}\frac{p^{k-1}(p-1)}{2}+\sum_{k=m+1}^{n}p^m\sum_{x\in p^{n-k+m}D_0^{(p^{k-m})}}\omega_N^{x}\nonumber\\
&=&1+\frac{p^m-1}{2}+p^m\sum_{k=m+1}^{n}\eta_0^{(p^{k-m})}\nonumber\\
&=&1+\frac{p^m-1}{2}+p^m\eta_0^{(p)}=\frac{p^m+1}{2}+p^m\eta_0^{(p)},\nonumber
\end{eqnarray}
The result follows. \ \ \ \ \ \ \ \ \ \ \ \ \ \ \ \ \ \ \ \ \ \ \ \ \ \ \ \ \ \ \ \ \ \ \ \ \ \ \ \ \ \ \ \ \ \ \ \ \ \ \ \ \ \ \ \ \ \ \ \ \ \ \ \ \ \ \ \ \ \ \ \ \ \ \ \ \ \ \ \ \ \  \ \ \ \ $\Box$

\begin{lemma}\label{num-theory}
Let $p$ be an odd prime, $n$ a positive integer, and $N=p^n$. Then, for $1\leq m\leq n-1$, we have $\mathrm{gcd}(2^{p^m}-1,\frac{2^N-1}{2^{p^m}-1})=\mathrm{gcd}(2^{p^m}-1,p^{n-m})$.
\end{lemma}
\noindent{\bf Proof.}
Since
\begin{eqnarray}
2^N-1&=&(2^{p^n}-1=2^{p^m}-1)(2^{p^m(p^{n-m}-1)}+2^{p^m(p^{n-m}-2)}+\cdots+2^{p^m}+1),\nonumber
\end{eqnarray}
then $\frac{2^{p^n}-1}{2^{p^m}-1}\equiv p^{n-m}\pmod{2^{p^m}-1}$, i.e.,
$$\mathrm{gcd}(2^{p^m}-1,\frac{2^N-1}{2^{p^m}-1})=\mathrm{gcd}(2^{p^m}-1,p^{n-m}).$$
\ \ \ \ \ \ \ \ \ \ \ \ \ \ \ \ \ \ \ \ \ \ \ \ \ \ \ \ \ \ \ \ \ \ \ \ \ \ \ \ \ \ \ \ \ \ \ \ \ \ \ \ \ \ \ \ \ \ \ \ \ \ \ \ \ \ \ \ \ \ \ \ \ \ \ \ \ \ \ \ \ \  \ \ \ \ \ \ \ $\Box$

\begin{theorem}\label{2-adic of order d}
Let $p$ be an odd prime and $n$ a positive integer. Suppose $\{s_i\}_{i=0}^{N-1}$ is the Ding-Helleseth sequence with period $N=p^n$ defined as in Eq. (\ref{sequence defi}). Then the 2-adic complexity $\phi_2(s)$ of $\{s_i\}_{i=0}^{N-1}$ is bounded by
\begin{equation}
\phi_2(s)\geq p^n-p^{n-1}-1>\frac{N+1}{2}.
\end{equation}
Specially, for $n=1,2$, the 2-adic complexity of $\{s_i\}_{i=0}^{N-1}$ is maximal.
\end{theorem}
\noindent{\bf Proof.}
Using Lemmas \ref{main method-2} and \ref{main method-3}, we have
\begin{eqnarray}
\mathrm{det}(A)&=&\prod_{a=0}^{N-1}S(\omega_N^a)\nonumber\\
&=&(\frac{p^n+1}{2})\prod_{m=0}^{n-1}\prod_{a\in p^mD_0^{(p^{n-m})}}S(\omega_N^a)\prod_{a\in p^mD_1^{(p^{n-m})}}S(\omega_N^a)\nonumber\\
&=&(\frac{p^n+1}{2})\prod_{m=0}^{n-1}(\frac{p^m+1}{2}+p^m\eta_0^{(p)})^{\frac{p^{n-m-1}(p-1)}{2}}(\frac{p^m+1}{2}+p^m\eta_1^{(p)})^{\frac{p^{n-m-1}(p-1)}{2}}\nonumber\\
&=&(\frac{p^n+1}{2})\prod_{m=0}^{n-1}\left((\frac{p^m+1}{2})^2+\frac{(p^m+1)p^m}{2}(\eta_0^{(p)}+\eta_1^{(p)})+p^{2m}\eta_0^{(p)}\eta_1^{(p)}\right)^{\frac{p^{n-m-1}(p-1)}{2}}.\nonumber
\end{eqnarray}
By Lemmas \ref{n equal one} and \ref{prod}, we obtain
\begin{eqnarray}
\mathrm{det}(A)=\left\{
\begin{array}{ll}
(\frac{p^n+1}{2})\prod_{m=0}^{n-1}(\frac{p^{2m+1}-1}{4})^{\frac{p^{n-m-1}(p-1)}{2}},\ \mathrm{if}\ p\equiv1\ (\mathrm{mod}\ 4), \\
(\frac{p^n+1}{2})\prod_{m=0}^{n-1}(\frac{p^{2m+1}+1}{4})^{\frac{p^{n-m-1}(p-1)}{2}},\ \mathrm{if}\ p\equiv3\ (\mathrm{mod}\ 4), \\
\end{array}
\right.
\end{eqnarray}
Now, suppose that $r$ is a prime factor of $2^{N}-1$ and that the multiplicative order of 2 modular $r$ is $\mathrm{Ord}_r(2)$. Then $\mathrm{Ord}_r(2)|N$, i.e., $\mathrm{Ord}_r(2)|p^n$. Since $p$ is a prime, then there exists an integer $1\leq t\leq n$ such that $\mathrm{Ord}_r(2)=p^t$. Next, we will prove $\mathrm{gcd}(\mathrm{det}(A),r)=1$ if $t=n$. We only prove this result for the case of $p\equiv1\ (\mathrm{mod}\ 4)$. For the case of $p\equiv3\ (\mathrm{mod}\ 4)$, it can be similarly proved. Suppose that  $r\mid\mathrm{det}(A)$. Then there exists an integer $0\leq m\leq n-1$ such that $r\mid \frac{p^{2m+1}-1}{4}$ or $r\mid \frac{p^{n}+1}{2}$. Note that $\mathrm{Ord}_r(2)=p^n$
and that $r\mid 2^{r-1}-1$, which implies $p^n\mid r-1$. Then $r\geq p^n+1$ and there exists a positive integer $s$ such that $r=sp^n+1$. It is obvious that $\frac{p^n+1}{2}<p^n+1$ and that $\frac{p^{2m+1}-1}{4}<p^n+1$ if $2m+1\leq n$, which implies that $\mathrm{gcd}(\frac{p^n+1}{2}, r)=1$ and that $\mathrm{gcd}(\frac{p^{2m+1}-1}{4},r)=1$ if $2m+1\leq n$. Therefore, we have $n<2m+1\leq 2n-1$ and there exists some positive integer $s_0$ such that $p^{2m+1}-1=rs_0=(sp^n+1)s_0$, i.e.,
$$(p^{2m+1-n}-1)p^n+(p^n-1)=ss_0p^n+s_0$$, which implies $s_0\equiv p^n-1\ (\mathrm{mod}\ p^n)$. But, on the other hand, we know that  $s_0=\frac{p^{2m+1}-1}{sp^n+1}\leq p^{n-1}$, a contradiction. Hence, we know that each prime factor of $\mathrm{gcd}(\mathrm{det}(A), 2^N-1)$ must be a prime factor of $2^{p^{n-1}}-1$. Furthermore, by Lemma \ref{num-theory}, $\mathrm{gcd}(2^{p^{n-1}}-1, \frac{2^N-1}{2^{p^{n-1}}-1})=\mathrm{gcd}(2^{p^{n-1}}-1,p)$. But we know that $\mathrm{gcd}(\mathrm{det}(A),p)=1$, which implies that
$\mathrm{gcd}(\mathrm{det}(A),\frac{2^N-1}{2^{p^{n-1}}-1})=1$. Thus we obtain
$$\phi_2(s)\geq \frac{2^N-1}{\mathrm{gcd}(\mathrm{det}(A),2^N-1)}\geq\frac{2^N-1}{2^{p^{n-1}}-1}\geq N-p^{n-1}-1$$

Specially, if $n=1,2$, then the sequences have been proved to be with maximal 2-adic complexity by Xiong et al. \cite{Xiong Hai} and Zeng et al. \cite{Zeng Xiangyong} respectively. The desire result follows.
\ \ \ \ \ \ \ \ \ \ \ \ \ \ \ \ \ \ \ \ \ \ \ \ \ \ \ \ \ \ \ \ \ \ \ \ \ \ \ \ \ \ \ \ \ \ \ \ \ \ \ \ \ \ \ \ \ \ \ \ \ \ \ \ \ \ \ \ \ \ \ \ \ \ \ \ \ \ \ \ \ \ \ \ \ \ \ \ \ \ \ \ \ \ \ $\Box$
\section{Summary and concluding remarks}
In this paper, we derive the gauss periods based on Ding-Helleseth generalized cyclotomic classes of order 2 from the prime power $p^n$. As an application, a lower bound on the 2-adic complexity of the Ding-Helleseth generalized cyclotomic sequence with length $p^n$ is determined. The result shows that the 2-adic complexity of these sequences is at least $p^n-p^{n-1}-1$, which is larger than the half of period $p^n$ and can resist the RAA for FCSRs.
\section*{Acknowledgement}
Parts of this work were written during a very pleasant visit of the first author to the Carleton University in
School of Mathematics and Statistics. She wishes to thank the hosts for their hospitality.

\end{document}